\pdfoutput=1
\documentclass{iopart}
\usepackage[utf8x]{inputenc}
\usepackage[T1]{fontenc}
\usepackage{breqn}
\usepackage{siunitx}
\usepackage{braket}
\usepackage[colorlinks, linkcolor=black, citecolor=black]{hyperref}
\usepackage[noabbrev]{cleveref}
\usepackage[usenames,dvipsnames]{color}
\usepackage{graphicx}
\usepackage{subfig}
\usepackage{textcomp}
\usepackage{bbm}

\newcommand{\gtwo}{$g_h^{(2)}(0)$}

\begin{document}
\title{Improving SPDC single-photon sources via extended heralding and feed-forward control}
\author{M Massaro, E Meyer-Scott, N Montaut, H Herrmann and C Silberhorn}
\ead{marcello.massaro@upb.com}

\address{Integrated Quantum Optics, University of Paderborn, Warburger Str. 100, D-33098 Paderborn, Germany}

\begin{abstract}
  Evolving photonic quantum technologies and applications require higher and higher rates
  of single photon generation. In parallel, it is required that these
  generated photons are kept spectrally pure for multi-photon experiments
  and that multi-photon noise be kept
  to a minimum. In spontaneous parametric down-conversion sources, these
  requirements are conflicting, because spectral filtering to increase spectral purity always means lowering
  the rate at which photons are generated, and increasing the pump power means
  increasing the multi-photon noise.
   In this paper, we present a scheme,
  called extended heralding,
  which aims to mitigate the reduction of single-photon generation rate under
  spectral filtering by removing cases where we detect light 
  in the rejection band of the heralding photon's filter.
   Our experiment shows that this allows for higher 
  single-photon generation rates with lower multi-photon noise than the standard approach
  of neglecting modes falling out of the filter bandwidth. We also 
  show that by using active feed-forward control based on this extended 
  heralding, it is possible to further improve the performance of the original  
  source by physically eliminating uncorrelated photons from the output stream.
\end{abstract}
\maketitle

\section{Introduction}\label{sec:introduction}

Many important tasks in optical quantum information processing, like boson sampling~\cite{Spring798,Tillmann:2013aa,Crespi:2013aa}, 
linear optics quantum computing~\cite{Carolan14082015,RUD17}, and quantum networking~\cite{Azuma:2015aa,Herbst17112015}, 
require the generation of single-photon states at high rates and with high purity.
Current methods for single-photon
generation include parametric downconversion (PDC) processes in non-linear
crystals and waveguides~\cite{evans_bright_2010,tanzilli_highly_2001},
four-wave mixing in optical fibers and waveguides~\cite{soller_high-performance_2011,fan_efficient_2005,1367-2630-13-6-065005}, and
quantum
dots~\cite{michler_quantum_2000,shields_semiconductor_2007,strauf_high-frequency_2007}
and color centers~\cite{alleaume_experimental_2004,gaebel_stable_2004,wu_room_2007} in solid state lattices.

Although all of these technologies are able to produce single
photons, the \textit{de facto} standard in applications is still PDC sources based on
non-linear crystals, due the proven technology behind their manufacturing
processes and simple opreration. They are also compact, able to produce photon
states of high spectral purity and indistinguishability~\cite{mosley_heralded_2008,Bruno:14} and easy to
package in integrated, room-temperature devices~\cite{Vergyris:2017aa,montaut_high-efficiency_2017}. 
These sources (called HSPS, or heralded
single photon sources) produce photons in pairs, giving the
user the opportunity to herald the presence of one photon by detecting its twin.

To obtain spectrally pure photons 
 suitable for interference with other photons,
two choices are possible: specifically engineering the HSPS itself in order to
ensure pure photons from the beginning~\cite{PhysRevA.64.063815}, or 
filtering a spectrally-correlated source~\cite{PhysRevLett.80.3891}.
Often, spectral filtering is the simpler way to obtain a pure, uncorrelated
spectrum. On the other hand, these being \textit{single} photon sources, we
want to keep the noise due to multi-photon components to a minimum. 
To avoid such noise, it is common to pump HSPSs at low power levels, such
that the probability of producing a single pair is small ($p<0.1$). While this
reduces the multi-photon component noise, it also reduces the number of ``useful
events'' per system-cycle (heralding rate), because most of the time no pairs will be produced.
The heralding rate drops even more when spectrally pure single photons are required,
since filtering in the heralding arm inevitably suppresses parts of the source
output. Increasing the pump power to recover the “lost” heralding rate will then
introduce substantial multi-photon noise, such that simply increasing the pump
power is not a viable solution. This is because the filter is applied only to
the heralding arm to maintain 
high heralding efficiency of the heralded photon~\cite{PhysRevA.95.061803},
and light which is outside the filter in the heralding arm does not lead to heralding events.
Nevertheless, it still exists in the heralded arm, leading to extra noise. Loosely, the 
heralding rate goes with $p_f$, the filtered photon production probability, while the 
noise in the heralding arm depends on $p~(>p_f)$, the original production probability
before filtering.

To improve the single photon generation rate of these sources while keeping the noise low,
multiple strategies can be used. A difficult but promising approach is multiplexing,
where multiple sources are combined together with feedforward switching to improve the
rate of single photon generation without increasing the noise. Common schemes are
spatial multiplexing~\cite{francis-jones_all-fiber_2016,meany_hybrid_2014},
time multiplexing~\cite{kaneda_time-multiplexed_2015} and frequency
multiplexing~\cite{joshi_frequency_2018,grimau_puigibert_heralded_2017}. 

While these are encouraging results, they usually come with a large hardware
overhead and additional setup complexity and losses when compared to single sources.
To avoid these drawbacks other methods
can be used to achieve better performance in certain metrics from a single SPDC source.
For example, a combination of high pump repetition rate
and spectral filtering has been used~\cite{ngah_ultra-fast_2015} to increase
the generation rate of single-photons. Alternatively, photon-number-resolving
detection can be used to eliminate higher order photon contributions, decreasing the multi-photon noise
~\cite{moseley_asymmetric-2018,PhysRevA.85.023829}. 

Note however, when the herald photon is filtered, many unheralded photons are sent to the heralded arm.
Some systems cannot rely on postprocessing alone to eliminate these unheralded photons,
 due to their sensitivity
on incoming photon flux (e.g. detection systems based on transition edge sensors~\cite{avella_self_2011}). In this case,
active feed-forward and electro-optical switches provide a solution, only opening the path to the heralded detector
after the herald photon has been detected~\cite{Brida:11,Brida2012An-extre,2040-8986-19-10-104005},
This approach uses off-the-shelf telecom equipment and thus relies on proven, cost-effective technology only.

In this paper, we present a scheme which ensures high spectral and photon-number
purity of the generated state by conditioning on heralding events both within and outside
the filter bandwidth, and add to this an active feed-forward
strategy that physically removes unwanted photons.  Our method does not
pollute the heralded arm of the PDC state with noise photons and constrains the
photon flux to a minimum for single-photon sensitive applications.
We compare our scheme with the case of standard passive spectral filtering and
measure the reduction in the noise due to our removal of higher order photon number contributions.
A measure of this noise is given by the heralded second
order correlation function (\gtwo) and we register a maximum improvement of
\SI{21}{\percent}, limited by losses in the heralding arm. 

\section{Theory and simulations}\label{sec:theory_simulations}

\begin{figure}[tb]
 \centering
 \includegraphics[width=0.7\textwidth]{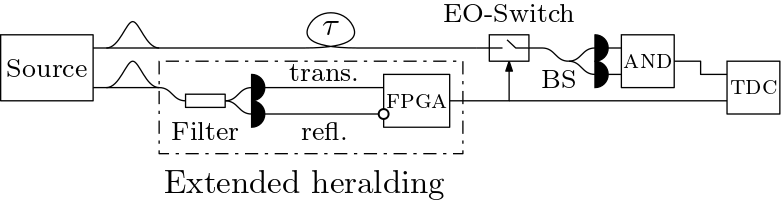}
 \caption{Representation of the proposed scheme. Light coming from both
 transmission and reflection port of a frequency filter is analysed by fast
 electronics (FPGA). The FPGA closes an electro-optic switch when a photon
 is present only in the transmission port. The statistics of the prepared
 state is then analysed with a Time-to-Digital Converter (TDC).
 }
 \label{fig:scheme}
\end{figure}

For the best heralding efficiency, often only the heralding photon is filtered. 
This can still project the heralded photon into a single spectral mode as long as the 
filter is tight enough~\cite{PhysRevA.95.061803}. A problem arises if the modes
which the filter removes in the 
heralding arm are still present in the heralded arm, leading to uncorrelated
noise.
In particular highly multi-mode PDC states with strong spectral correlations,
which are typical for standard waveguide sources, require significant filtering
to achieve spectral purity and are strongly affected by this pollution.
 To counteract this shortcoming we introduce the concept of \textit{extended
heralding} (\cref{fig:scheme}). Here we exploit the additional information gained by
detecting not only the transmitted light of filter in the heralding arm, but
also the rejected part of the spectrum which is normally discarded. We use the 
additional information present in the reflected heralding mode to improve the 
photon statistics of the heralded mode, namely by discounting events where
a photon is detected in both transmitted and reflected ports,
 reducing multi-photon events. We follow the description of
Ref.~\cite{PhysRevA.85.023829} with the addition of filtering and extended heralding,
comparing the fidelity of the heralded state to a single photon in a single 
spectral
mode, versus the probability of heralding. Here we take the fidelity of the
heralded state before any losses, which is thus an upper limit to the fidelity
achievable in practice.

\subsection{Effect of filtering on fidelity}\label{sub:filtering_fidelity}
We begin with a PDC state with a certain distribution of spectral modes given
by the Schmidt decomposition of the joint spectral amplitude of the photon
pairs. The state in terms of these broadband modes is~\cite{PhysRevA.85.023829}
\begin{eqnarray}
  \ket{\psi}&=\bigotimes_{k=0}^\infty\mathrm{sech}(q_k)\sum_{n=0}^\infty 
\tanh^n(q_k)\ket{n_k^{(s)},n_k^{(i)}}.
\end{eqnarray}
This means, for each spectral-temporal Schmidt mode $k$ in the tensor product, 
there is a
sum in photon number from 0 to $\infty$, with a thermal distribution. The
squeezing parameters are defined as $q_k = B\lambda_k$, where $B$ is an
overall pump power factor, and $\lambda_k$ are the eigenvalues of the Schmidt
decomposition of the joint spectral amplitude. In practice the tensor product
and sum need not be carried to infinity; we use a maximum of 20 spectral modes
and 6 photons in simulation.

The spectral filter in the heralding (signal) arm rearranges the spectral
modes. It suffices in most cases to take new (pseudo-)Schmidt
decompositions of the joint spectral amplitudes transmitted and reflected
by the filter~\cite{PhysRevA.90.023823,1367-2630-12-6-063001},
without renormalizing. Then the state, with components now labelled $t$ for 
transmitted through the filter and $r$ for reflected, is
\begin{eqnarray}
  \ket{\psi}&=\bigotimes_{k_t=0}^\infty\mathrm{sech}(q_{k_t})\sum_{n=0}^\infty 
\tanh^n(q_{k_t})\ket{n_{k_t}^{(s)},n_{k_t}^{(i)}}\\\nonumber
            &\bigotimes_{k_r=0}^\infty\mathrm{sech}(q_{k_r})\sum_{n=0}^\infty 
\tanh^n(q_{k_r})\ket{n_{k_r}^{(s)},n_{k_r}^{(i)}},
\end{eqnarray}
where now the squeezing parameters are $q_{k_t} = B_t\lambda_{k_t}$ and
$q_{k_r} = B_r\lambda_{k_r}$, and $B_r$ and $B_t$ come from the relative
intensities of the transmitted and reflected modes.

We can sort the terms according to photon number as
\begin{dmath}
  \ket{\psi}=\prod_{k_t}\mathrm{sech} (q_{k_t})\left(\ket{0_t}+\sum_{k_t} 
\tanh(q_{k_t})\ket{1_{k_t}^{(s)};1_{k_t}^{(i)}}+\sum_{k_t\le k_t^\prime} 
\tanh(q_{k_t})\tanh(q_{k_t^\prime})\ket{1_{k_t}^{(s)},1_{k_t^\prime}^{(s)};1_{
k_t}^{(i)},1_{k_t^\prime}^{(i)}}\dots\right)\\\nonumber
\otimes\prod_{k_r}\mathrm{sech}(q_{k_r})\left(\ket{0_r}+\sum_{k_r} 
\tanh(q_{k_r})\ket{1_{k_r}^{(s)};1_{k_r}^{(i)}}+\sum_{k_r\le k_r^\prime} 
\tanh(q_{k_r})\tanh(q_{k_r^\prime})\ket{1_{k_r}^{(s)},1_{k_r^\prime}^{(s)};1_{
k_r}^{(i)},1_{k_r^\prime}^{(i)}}\dots\right),
\end{dmath}
and continuing with higher number terms. Then we apply a detector (insensitive 
to the spectral-termporal modes) on the
heralding (transmitted) mode, given by the heralding projector
\begin{dmath}
  \hat{\Pi}_t=c_0\ket{0}\bra{0} + c_1\sum_{k_t}\ket{1_{k_t}}\bra{1_{k_t}} + 
c_2\sum_{k_t\le k_t^\prime}\ket{1_{k_t},1_{k_t^\prime}}
\bra{1_{k_t},1_{k_t^\prime}}+\dots,
\end{dmath}
where $c_{n_t} = 1-\left(1-\eta_t\right)^{n_t}$ is the click probability given
$n_t$ photons for detection efficiency $\eta_t$. The extended heralding
(reflected mode) projector is then $\hat{\Pi}_r=\mathbbm{1}-\hat{\Pi}_t$. This
is equivalent to $\hat{\Pi}_t$, with the detection probabilities $c_{n_t}$
replaced by the probabilities of no click, $c_{n_r} =  (1-\eta_r)^{n_r}$. Dark
counts can be added to either detector with a constant term in the $c_n$.

Projecting the transmitted and reflected modes with their respective heralding
and extended heralding detectors returns the (normalized) heralded 
signal state
\begin{eqnarray}\label{eq.heraldedstate}
\rho_s=&\frac{\prod_{k_t}\mathrm{sech^2}(q_{k_t})}{p_{herald}}\left(c_{0_t}\ket{
0_t}\bra{0_t} + c_{1_t} \sum_{k_t} \tanh^2(q_{k_t})\ket{1_{k_t}}\bra{1_{k_t}}\right.\\\nonumber
        &\left.+c_{2_t}\sum_{k_t\le k_t^\prime}\tanh^2(q_{k_t})\tanh^2(q_{k_t^\prime})\ket{1_{k_t},1_{k_t^\prime}}\bra{1_{k_t},1_{k_t^\prime}}+\dots\right)\\\nonumber
        &\otimes\frac{\prod_{k_r}\mathrm{sech^2}(q_{k_r})}{p_{ext}}\left(c_{0_r}\ket{0_r}\bra{0_r}+ c_{1_r} \sum_{k_r}\tanh^2(q_{k_r})\ket{1_{k_r}}\bra{1_{k_r}}\right.\\\nonumber
        &\left.+c_{2_r}\sum_{k_r\le k_r^\prime}\tanh^2(q_{k_r})\tanh^2(q_{k_r^\prime})\ket{1_{k_r},1_{k_r^\prime}}\bra{1_{k_r},1_{k_r^\prime}}+\dots\right).
\end{eqnarray}
The probabilities of heralding and extended heralding (i.e.\ getting a click in
the transmitted arm, and {\em no} click in the reflected arm, respectively) are
\begin{eqnarray}
  \fl p_{herald}=\prod_{k_t}\mathrm{sech^2}(q_{k_t})\left(c_{0_t} + c_{1_t}  
\sum_{k_t} \tanh^2(q_{k_t})\right.\\\nonumber
  \left.+c_{2_t}\sum_{k_t\le k_t^\prime}\tanh^2(q_{k_t})\tanh^2(q_{k_t^\prime})+\dots\right)
\end{eqnarray}
and
\begin{eqnarray}
\fl p_{ext}=\prod_{k_r}\mathrm{sech^2}(q_{k_r})\left(c_{0_r} + c_{1_r}  \sum_{k_r}\tanh^2(q_{k_r})\right.\\\nonumber
\left.+c_{2_r}\sum_{k_r\le k_r^\prime}\tanh^2(q_{k_r})\tanh^2(q_{k_r^\prime})+\dots\right).
\end{eqnarray}

\begin{figure}[!tb]
\centering
	\includegraphics[scale=.7]{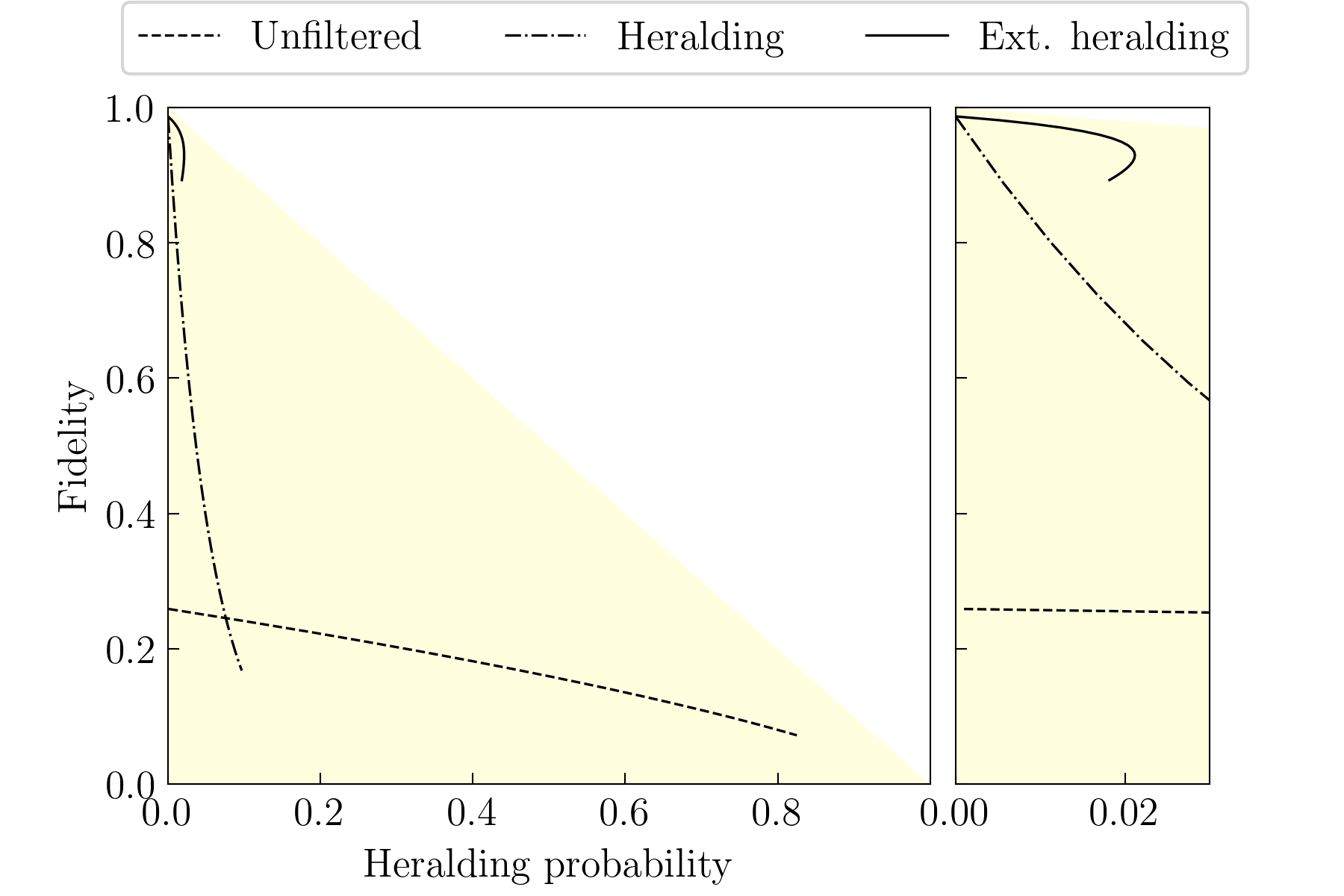}
	\caption{
	Left plot: theoretical fidelity versus heralding probability, with and 
without spectral filtering and extended heralding. Each curve is 
produced 
by varying the initial (before filtering) mean photon number over 
$[10^{-2},2]$. After filtering and extended heralding, the state reaches and 
remains in the high-fidelity region for a range of pump powers. Right plot:
Detail of the aforementioned range in which the fidelity is consistently higher
than both unfiltered and heralded case.}
	\label{fig:fidelity_theory}
\end{figure}

Finally, the fidelity of the heralded state to a single photon in the first 
spectral mode and vacuum in all other modes is
\begin{eqnarray}\label{eq.fidelity}
F &= \bra{1_{0_t}}\otimes\bra{0_r}\rho\ket{1_{0_t}}\otimes\ket{0_r} \\\nonumber
&=\frac{\prod_{k_t}\mathrm{sech^2}(q_{k_t})}{p_{herald}}\frac{\prod_{k_r}\mathrm
{sech^2}(q_{k_r})}{p_{ext}}c_{1_t}c_{0_r}\tanh^2(q_{0_t}).
\end{eqnarray}
This is the fidelity of the heralded single photon before it is subjected to 
losses,
upper bounding the possible performance of the source.
This fidelity is plotted in \cref{fig:fidelity_theory}, which shows the
progression from an unfiltered, spectrally multimode state (with JSI matching
our experiment), to filtered but contaminated with multiphoton events, to
finally a high-fidelity state. An ideal photon source has simultaneously high
heralding probability and fidelity, but without multiplexing heralded single
photon sources are limited to the yellow region. Of course, one still wants
to operate the sources as close to the upper right corner as possible. Here
the unfiltered state shows consistently low fidelity, which can be increased
by filtering, in our case to \SI{50}{\giga\hertz} bandwidth. Without losses, the
extended heralding case maintains high fidelity for significantly 
higher, as shown in the right plot of \cref{fig:fidelity_theory},
heralding probability than the standard filtered case, approaching the
theoretical limit much more closely. 



It is unfortunately not possible to access this fidelity experimentally due to
the difficulty of projecting on the vacuum and single photons in single
spectral modes with realistic losses. The source quality can still be
accessed, however, via the spectral purity $P$ and heralded \gtwo.
Fidelity $F=1$ corresponds to $P=1$ and $g_h^{(2)}(0)=0$. Here the purity is
controlled by spectral filtering, which increases the \gtwo\ from pollution
from other spectral modes, which is subsequently reduced by heralding plus 
feed-forward.

We now make a few approximations to express the fidelity \cref{eq.fidelity} 
in terms
of these experimentally accessible quantities. First we assume all $q_{k_t}$ and 
$q_{k_r}$ 
are small such that $\mathrm{tanh}(q_k)\approx q_k$ and 
$\mathrm{sech}(q_k)\approx 1$.
Then we neglect filtering (droping the subscripts $t$), resulting in a fidelity 
\begin{equation}
F\approx \frac{c_1q_0^2}{p_{herald}}.
\end{equation}
We then assume low overall generation probability and no dark counts, such that 
we can truncate 
$p_{herald}$ to second order, giving
\begin{equation}
p_{herald}\approx c_{1}  \sum_{k} (q_{k})^2+c_{2}\sum_{k\le k^\prime} 
(q_{k})^2(q_{k^\prime})^2.
\end{equation}
We expand the fidelity in a Taylor series about 0 in the second term of 
$p_{herald}$ ($1/(a+x)\approx 1/a - x/a^2)$, giving
\begin{equation}
F\approx \frac{q_0^2}{\sum_{k} (q_{k})^2} \left(1-\frac{c_{2}\sum_{k\le 
k^\prime} 
(q_{k})^2(q_{k^\prime})^2}{c_1\sum_{k} (q_{k})^2}\right)
\end{equation}

The spectral purity is given in~\cite{1367-2630-13-3-033027} as
\begin{equation}
P \approx \frac{\sum_{k} (q_{k})^4}{\left(\sum_{k} (q_{k})^2\right)^2} \approx 
\frac{q_0^4}{\left(\sum_{k} (q_{k})^2\right)^2},
\end{equation}
where we assumed that the first Schmidt mode dominates, i.e. $q_0^4\gg 
q_{k>0}^4$. Then we can identify this purity with the square of the first term 
in our approximate fidelity, resulting in 
\begin{equation}
F\approx \sqrt{P}\left(1-\frac{c_{2}\sum_{k\le k^\prime} 
(q_{k})^2(q_{k^\prime})^2}{c_1\sum_{k} (q_{k})^2}\right).
\end{equation}

We similarly approximate and truncate the heralded state 
(\cref{eq.heraldedstate}) to give
\begin{equation}
\fl\rho_s\approx \frac{1}{p_{herald}}\left(
         c_{1} \sum_{k} (q_{k})^2\ket{1_{k}}\bra{1_{k}}
        +c_{2}\sum_{k\le k^\prime} 
(q_{k})^2(q_{k^\prime})^2\ket{1_{k},1_{k^\prime}}\bra{1_{k},1_{k^\prime}}
\right).
\end{equation}
The heralded \gtwo{} is given for multimode states~\cite{1367-2630-13-3-033027} 
with broadband mode operators $A_k$ by 
\begin{equation}
g_h^{(2)}(0) = \frac{\left\langle\left(\sum_{j,m}A_j^\dagger A_m^\dagger 
A_jA_m\right)\right\rangle}{\left\langle \sum_j A_j^\dagger A_j\right\rangle^2},
\end{equation}
which when substituting $\rho_s$ results in
\begin{equation}
g_h^{(2)}(0) \approx \frac{2c_{2}p_{herald}\sum_{k\le k^\prime} 
(q_{k})^2(q_{k^\prime})^2}{\left(c_1\sum_{k} (q_{k})^2\right)^2}.
\end{equation}

We then keep $p_{herald}$ only to first order, allowing to identify the second 
term of our approximate fidelity with \gtwo$/2$, resulting in 
\begin{equation}\label{eq.Fapprox}
F \approx \sqrt{P}\left(1-\frac{g_h^{(2)}(0)}{2}\right).
\end{equation}
For the filtered case, with or without extended heralding, see the 
derivation in the Appendix, which shows \cref{eq.Fapprox} still holds. That 
is, the purity and \gtwo{} capture the relevant features (to first order) to 
measure the lossless heralded single-photon fidelity in cases of filtering, no 
filtering, and extended heralding.

\subsection{Simulations}\label{sub:simulations}
To characterize the efficacy of extended heralding in
the presence of realistic spectral mode distributions, losses, and
higher-order photon states, we performed numerical simulations using
QuTiP~\cite{Johansson:2012aa}. We calculate the two-photon joint spectral
amplitude of our photon pair source, perform a Schmidt decomposition to find
the relative strengths of the involved squeezers, then normalize the overall
pump power to give the appropriate total mean photon pair number. Next
idealized (square, lossless) filters are applied to the heralding arm, and
detector operations are applied to the transmitted and reflected arms, and
we calculate the fidelity from \eref{eq.fidelity}. We also analyze the
spectral purity and photon statistics via the \gtwo\ in order to compare with
experiment. Finally we introduce the \textit{heralded single-photon fitness} 
$F_{HS}$,
so named because it captures two important aspects of a heralded single photon
source: the presence of one photon in a single spectral mode upon heralding, and 
the absence of
photons without heralding. The former is improved here by extended heralding,
 and the latter by using feedforward
to control a physical switch on the heralded mode. The contributions of these
terms are weighted by their likelihood. We
define the heralded single-photon fitness as 
\begin{eqnarray}
	\label{eq:source_fitness}
F_{HS} &= p_{herald}F + (1-p_{herald})P_{noclick}\\\nonumber
&\approx p_{herald} \sqrt{P} \left(1-\frac{g_h^{(2)}(0)}{2}\right) + 
(1-p_{herald})P_{noclick}
\end{eqnarray}
where $P_{noclick}$ is the probability of getting no detection in the heralded 
mode given that there was no heralding signal. This probability is also 
scaled to take into account lossess in the setup, i.e.
\begin{equation}
 P_{noclick} = 1-\frac{P_{click}}{\eta}
\end{equation}
where $\eta$ is the heralded photon's Klyshko efficiency, which gives the probability of
producing no photon inside the source.

\section{Experimental setup}\label{sec:experimental_setup}
\begin{figure}[!htb]
	\centering
	\includegraphics[width=\textwidth]{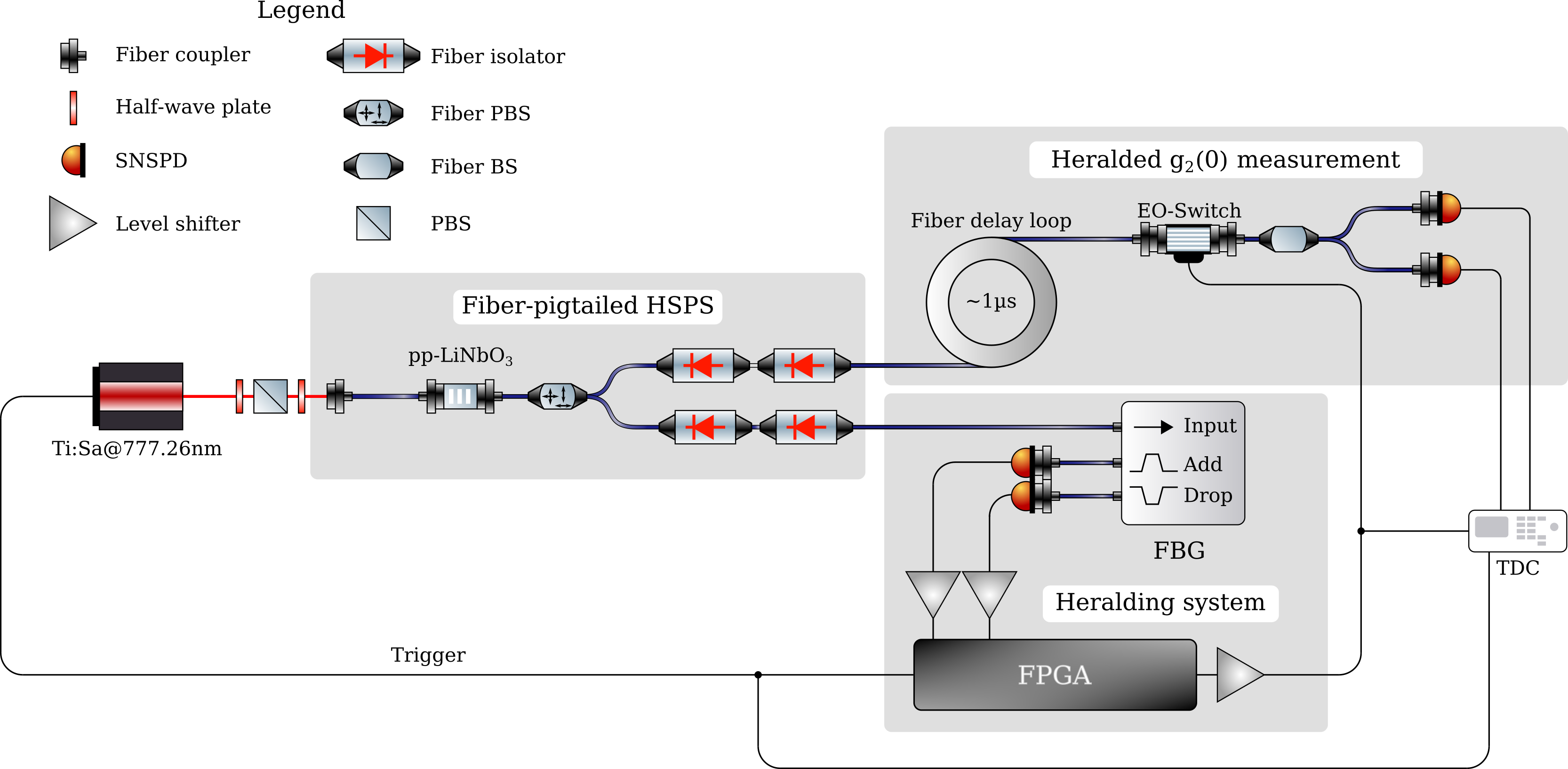}
	\caption{
        Experimental setup. Black thin lines represent electrical
	  connections, while thick blue lines are single-mode fibers. The
	  electro-optic switch (EO-Switch) is a 2x2 device whose unused
	  ports are not depicted. The Time-To-Digital converter is connected
	to a computer for data storage (not depicted).}
	\label{fig:setup}
\end{figure}

The experimental setup is shown in \cref{fig:setup}.
We use a type-II SPDC source based on a periodically-poled titanium-indiffused 
lithium-niobate
waveguide. The source is fiber-pigtailed and is engineered to produce single
photon pairs at \SI{1560}{\nano\meter} when pumped at \SI{780}{\nano\meter} and
kept at room temperature~\cite{montaut_high-efficiency_2017}. However, in this
experiment the source is kept at around \SI{50}{\celsius}, to shift
the degeneracy point of the source into a range that fits the window of the
fiber Bragg-Grating (FBG) filter used for heralding.

A pulsed laser system (\textit{Spectra Physics Tsunami}) with a 
\SI{2}{\pico\second}
bandwith and centered at \SI{777.24}{\nano\meter} with a repetition rate of
\SI{500}{\kilo\hertz} pumps the SPDC and acts as the system clock. In order 
to control power and polarization of the pump,
a half-wave plate, a polarizing beam-splitter, and a second half-wave plate,
are placed just before light is coupled into the SPDC source.
The pump is then coupled into a polarization maintaining fiber directly 
pigtailed
to one of the end facets of the lithium-niobate chip. Photon pairs are then collected at the
output end facet by another polarization maintaining fiber. The output fiber is
fusion-spliced to a fiber polarizing beam-splitter to separate the two outputs
of the source (typically called ``signal'' and ``idler''). The outputs of the 
fiber
PBS are then also spliced to two fiber isolators per arm to suppress the
residual pump light. In our case, the signal arm is the one being filtered
and the idler is the arm being analysed.

The signal arm is connected to an FBG filter (\textit{AOS Manual FBG}) centered
at \SI{1554.5}{\nano\meter} and with \SI{0.25}{\nano\meter}
bandwith. The two outputs of the filter correspond to the selected portion
of the spectrum we want to use for heralding (``add'' or transmitted) and the 
extended-heralding one (``drop'' or reflected).

To implement the schemes described above, we monitor
both outputs of the filter and apply two different heralding
criteria. In the first case, which we will use as benchmark and
label simply ``heralding'', the heralding signal is taken to be a click in the
transmitted arm of the filter. In the second case, labeled extended heralding, 
the heralding
signal is a combination of two events: a click in the transmitted arm and no 
clicks
in the reflected arm. Additionally, in this second case the extended heralding signal
is used to close an electro-optic switch. This means that we are physically
removing unheralded photons from the system instead of simply discarding
such events during data-analysis, a feature which proves crucial in
light sensitive applications.

In both cases, the signal arm photons are detected with an superconducting nanowire 
single-photon detector (SNSPD, \textit{PhotonSpot}). The generated outputs are amplified to TTL levels
and redirected to a field programmable gate-array (FPGA, \textit{Xilinx Spartan 
6})
which produces the final heralding signal, depending on the condition set.
This signal is then fed to a time-to-digital converter (TDC, 
\textit{AIT TTM8000}) for analysis.

The idler arm is routed to a fiber loop which introduces a delay 
of approximately
\SI{1}{\micro\second}. This gives enough time for the electronics to generate the
appropriate heralding signal as described above. This then is used to close an
electro-optic switch, which is normally open (i.e. blocking transmission of the light). The output of the switch
is then coupled to a non-polarizing beam-splitter, whose outputs are finally
connected to two SNSPDs.
Their outputs are analysed by the TDC, which registers the timestamps of each detector click
and saves them onto a computer.

\section{Analysis and discussion}\label{sec:analysis_and_discussion}

\begin{figure}[tb]
	\centering
	\subfloat[][\gtwo{} comparison between normal and extended heralding.]{
	\includegraphics[width=0.45\textwidth]{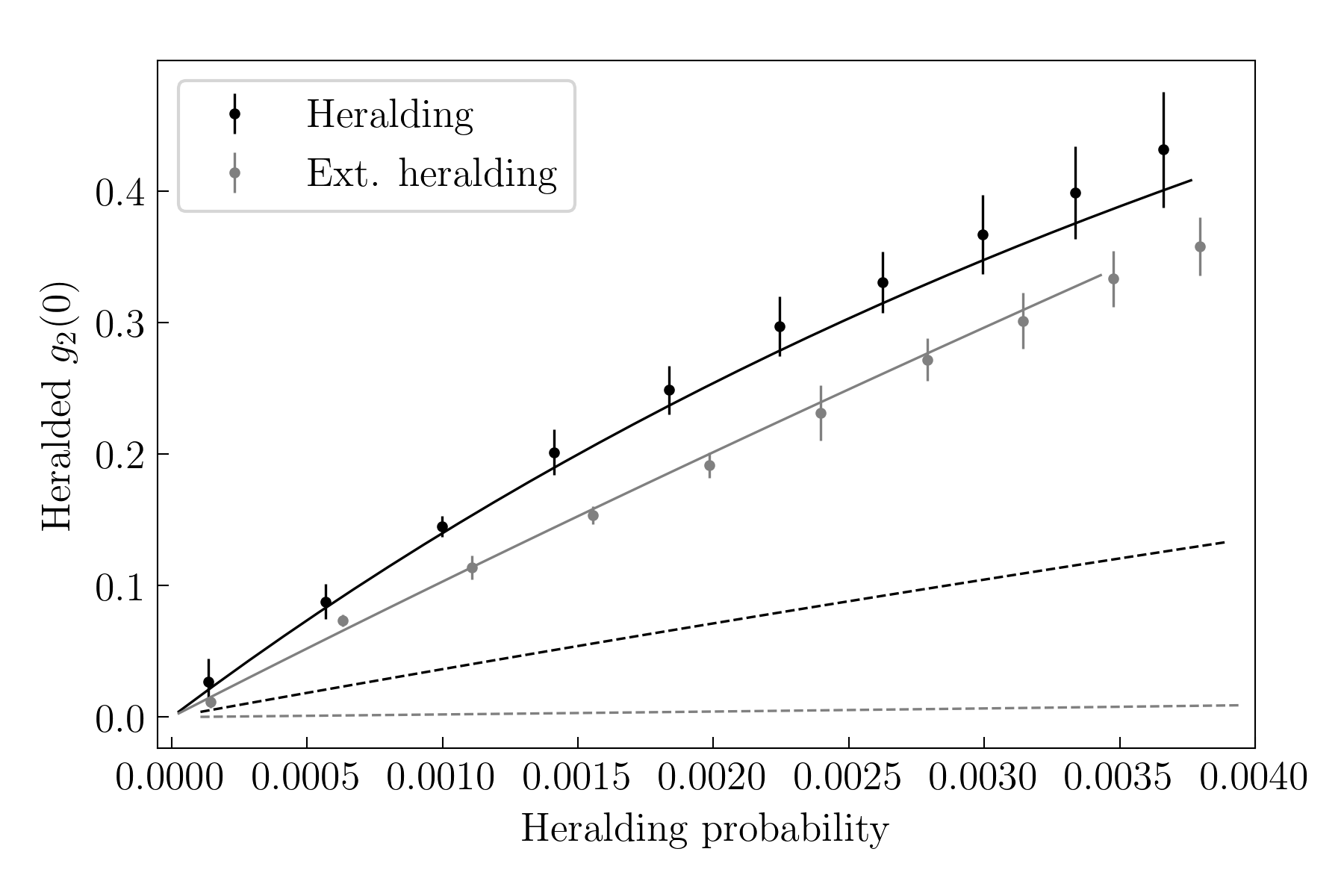}
	\label{fig:data}}\quad
    \subfloat[][Fidelity to the $\ket{1}$ state in a single 
        spectral mode.]{
	      	\includegraphics[width=0.45\textwidth]{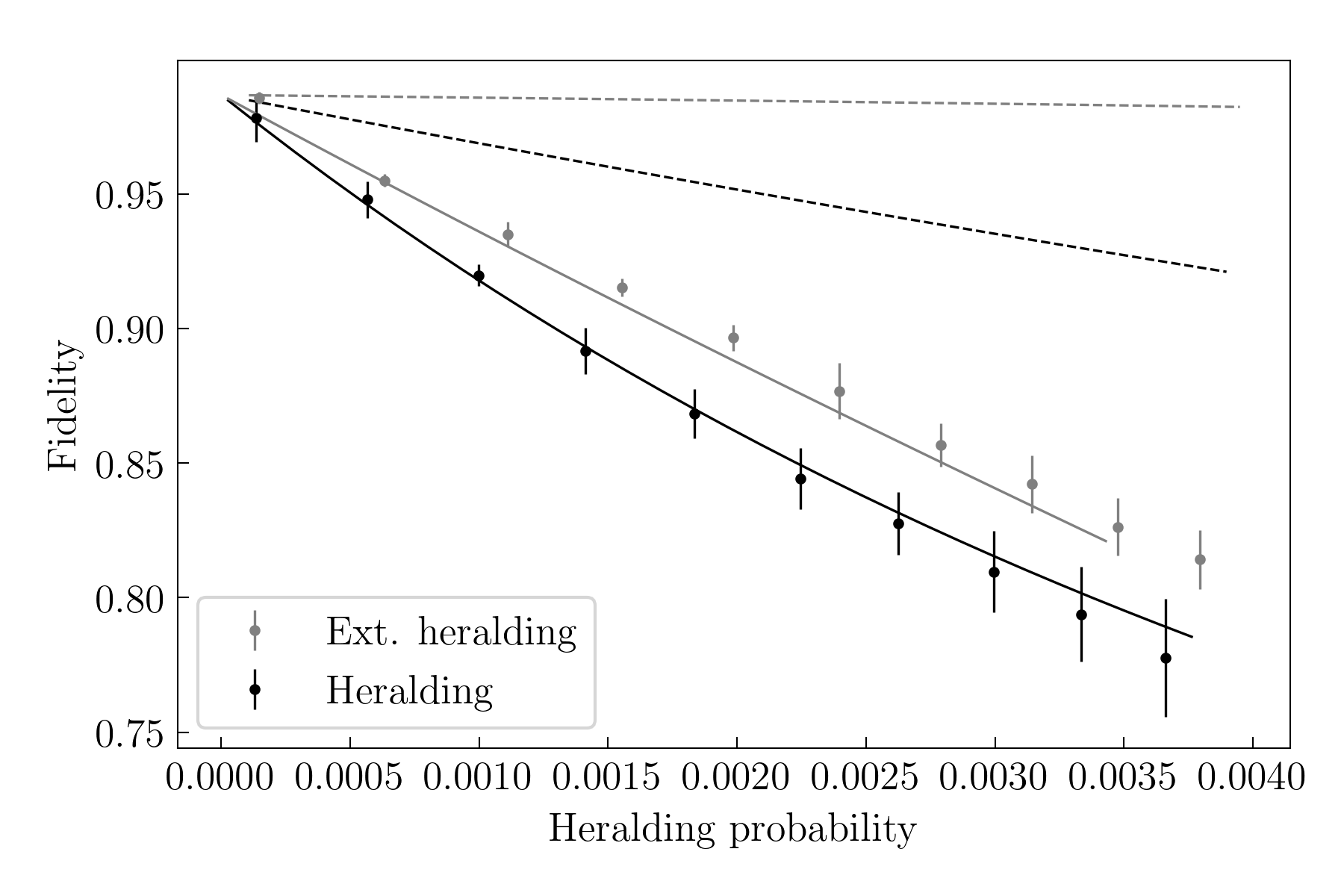}
          \label{fig:fidelity_comp}
        }
    \caption{Effect of normal and extended heralding. The two figures a) and b)
    represent equations \ref{eq:gtwo_exp} and \ref{eq.Fapprox}, respectively. Error bars are
    statistical errors with 1-sigma confidence interval calculated
    over ten repetitions of the experiment, each with an integration time of
    120 seconds. Solid lines represent
    our theoretical models, taking into
	account the experimental parameters. Dashed lines represent the same
	model in the case of lossless components throughout the setup.}
	\label{fig:results}
\end{figure}

As stated before, filtering is the simplest way to improve the purity of
a source with a correlated spectrum. This comes at a cost, namely a reduction in 
the
generation rate of single-photons. To counteract this effect, the natural
answer would be to increase the pump power used. This has the unwanted
side-effect to also increase the multi-photon components of the final state.
To show this behaviour, we record the \gtwo{} value at different power levels
when using the heralding signal from the transmitted FBG port.
We show the data as a function of heralding probability as only the photons
which have been heralded are usable. As summarized by
\cref{fig:data}, the \gtwo{} captures the undesired increase
of the multiphoton-component contribution as the heralding probability
increases. The \gtwo{} value is calculated
according to
\begin{equation}
	\label{eq:gtwo_exp}
	g_h^{(2)}(0) = \frac{C}{S_1S_2}H,
\end{equation}
where $C$ is the number of heralded coincidences at the end of the
non-polarizing beam-splitter, $S_1$ and $S_2$ are the heralded
counts at each output, respectively, and $H$ is the number of
heralding signals in the experiment.

If we now take into account the
reflected port of the FBG filter to generate the extended heralding signal,
we can see that, for the same value of heralding probability, we have
decreased the \gtwo{} value, indicating that we effectively mitigated the
spurious contributions of multiple pairs.  Additionally, the data
is in good agreement with out theoretical model, which allows us
to extrapolate the best possible improvement when sourcing components with
lower losses. 

\Cref{fig:data} and \ref{fig:fidelity_comp} show the measured 
values of the \gtwo{} and the fidelity,
with the respective theoretical models. We can see that in each case theory
and experiment are in good agreement, particularly for the solid lines, which 
directly implement in simulation the approximations made in the approximate 
fidelity \cref{eq.Fapprox}. An additional consideration is that this scheme is more
effective the harder the source is pumped, e.g. reaching a maximum improvement
in \gtwo{} of \SI{21}{\percent}, or for the same \gtwo{}, an improvement in the
count rate of 1.2 times.

The performance of extended heralding is mainly limited by the losses in the
experimental setup, as can be seen in the significant difference in the simulated
curves with and without losses. Fiber-to-fiber connections, the FBG itself and
the network connecting the source to the detection system all amount to
an estimated \SI{30}{\percent} total transmission. Sourcing better components
could increase the improvement obtained with this scheme over simple
heralding to a more than \SI{80}{\percent} reduction in the \gtwo{}.

\begin{figure}[t]
	\centering
    \includegraphics[scale=.7]{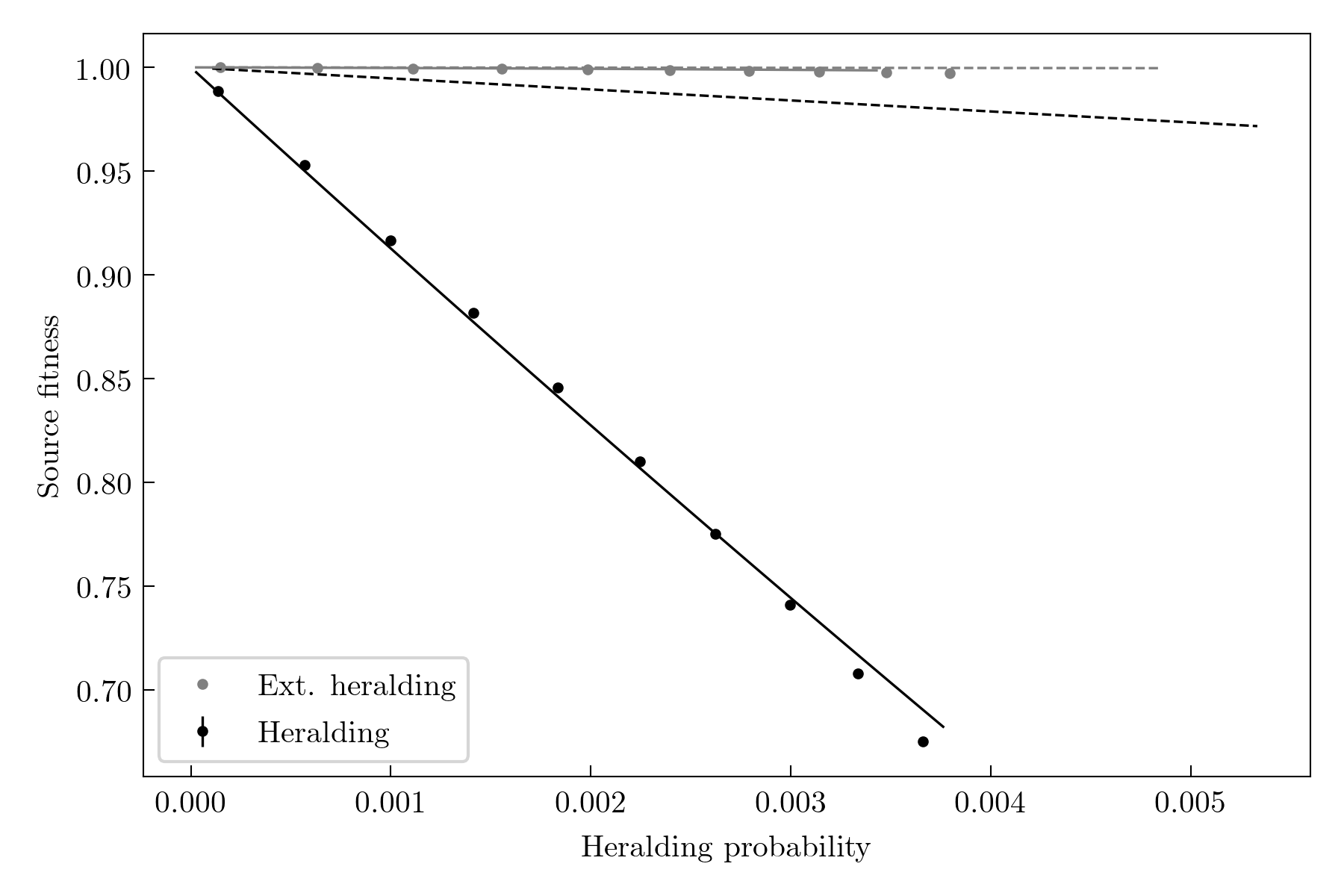}
    \caption{Source ``fitness'' as single photon source
    (\cref{eq:source_fitness}). As above, solid lines represent
    our theoretical models, and dashed lines represent the same
	model with lossless components.}
    \label{fig:source_fitness}
\end{figure}

In the previous two measurements, it is not necessary to implement feed-forward, as
the same result can be achieved with post-selection.
 However, active gating is of importance when the total light flux
reaching the experimental setup must be kept to a minimum, as it allows
only correctly heralded photons to pass. 
The source fitness parameter $F_{HS}$ as introduced in \cref{eq:source_fitness} 
directly captures this improvement, which cannot be achieved
with post-selection. In contrast to the improvement in fidelity, here the increase in photon fitness
is more significant for higher losses, as more losses in the heralding arm
mean more unheralded events make it to the detectors without feed-forward. As seen in \cref{fig:source_fitness},
the source fitness after extended heralding and feed-forward is nearly perfect, with a maximum improvement of \SI{53}{\percent}.

Another parameter used to characterize such active sources is
the output noise factor (ONF)~\cite{Brida:11}. Given an heralding probability of
0.0037, optical switch on-time of \SI{200}{\nano\second} and optical switch
extinction ratio of \SI{20}{\deci\bel}, we calculated a ONF of
\SI{2.4 +- 2.0}{\percent}.

\section{Conclusion}\label{sec:conclusion}
In this paper we have introduced and implemented a scheme called extended heralding, aimed
at improving the standard filtering used to increase the spectral purity of photon pair sources with 
a correlated joint spectrum. A significant improvement is found when  compared to a 
passive filter, especially when care is taken to minimise losses throughout the 
setup. This scheme is also easy to implement on top of an existing HSPS, requiring
no modification of the existing setup. 
Finally we have demonstrated a significant reduction in unwanted incident light through the use of active
feed-forward, which is important in practical light-sensitive scenarios.

\ack\label{sec:ack}
The authors thank Johannes Tiedau for helpful discussions, and acknowledge support from the European Commission, European Research Council (ERC) (725366 QuPoPCoRN). EMS acknowledges support from the Natural Sciences and Engineering Research Council of Canada. 

\section*{Appendix}
Here we derive again the approximate fidelity in the filtering and extended heralidng case, showing it gives the same result as \cref{eq.Fapprox}.  Again we assume all $q_{k_t}$ and $q_{k_r}$ 
are small such that $\mathrm{tanh}(q_k)\approx q_k$ and  $\mathrm{sech}(q_k)\approx 1$.
Then the fidelity is 
\begin{equation}
F\approx \frac{c_{1_t}c_{0_r}q_{0_t}^2}{p_{herald}p_{ext}}.
\end{equation}
We then assume low overall generation probability and no dark counts, such that we can truncate 
$p_{herald}$ to second order, giving
\begin{equation}
p_{herald}\approx c_{1_t}  \sum_{k_t} (q_{k_t})^2+c_{2_t}\sum_{k_t\le k_t^\prime} 
(q_{k_t})^2(q_{k_t^\prime})^2,
\end{equation}
and $p_{ext}$ to first order as
\begin{equation}
p_{ext}\approx c_{0_r} + c_{1_r}  \sum_{k_r} (q_{k_r})^2.
\end{equation}

We also neglect the third-order term in the product $p_{herald}p_{ext}$, giving fidelity
\begin{eqnarray}
\fl F\approx c_{1_t}c_{0_r}q_{0_t}^2\left(c_{1_t}c_{0_r}\sum_{k_t} (q_{k_t})^2\right.\\\nonumber
\left.+c_{2_t}c_{0_r}\sum_{k_t\le k_t^\prime}(q_{k_t})^2(q_{k_t^\prime})^2 + c_{1_t} c_{1_r}\sum_{k_t} (q_{k_t})^2 \sum_{k_r} (q_{k_r})^2\right)^{-1} .
\end{eqnarray}
We then expand the fidelity in a Taylor series about 0 in the second two terms of the denominator ($1/(a+x)\approx 1/a - x/a^2)$, giving
\begin{equation}
\fl F\approx \frac{q_0^2}{\sum_{k_t} (q_{k_t})^2} \left(1-\frac{c_{2_t}c_{0_r}\sum_{k_t\le k_t^\prime} 
(q_{k_t})^2(q_{k_t^\prime})^2}{c_{1_t}c_{0_r}\sum_{k_t} (q_{k_t})^2}-\frac{ c_{1_t} c_{1_r}  \sum_{k_t} (q_{k_t})^2 \sum_{k_r} (q_{k_r})^2}{c_{1_t}c_{0_r}\sum_{k_t} (q_{k_t})^2}\right)
\end{equation}

The spectral purity is the same as before~\cite{1367-2630-13-3-033027}
\begin{eqnarray}
P \approx \frac{\sum_{k_t} (q_{k_t})^4}{\left(\sum_{k_t} (q_{k_t})^2\right)^2} \approx \frac{q_0^4}{\left(\sum_{k_t} (q_{k_t})^2\right)^2}.
\end{eqnarray}

We now include the reflected modes in the truncated heralded state (\cref{eq.heraldedstate}) to give
\begin{eqnarray}
\fl \rho_s\approx \frac{1}{p_{herald}}\left(
         c_{1_t} \sum_{k_t} (q_{k_t})^2\ket{1_{k_t}}\bra{1_{k_t}}
        +c_{2_t}\sum_{k_t\le k_t^\prime} (q_{k_t})^2(q_{k_t^\prime})^2\ket{1_{k_t},1_{k_t^\prime}}\bra{1_{k_r},1_{k_r^\prime}}\right)\\\nonumber
        \otimes \frac{1}{p_{ext}}\left(c_{0_r}\ket{0}\bra{0}
        + c_{1_r} \sum_{k_r} (q_{k_r})^2\ket{1_{k_r}}\bra{1_{k_r}}\right).
\end{eqnarray}
The heralded \gtwo\ is given for multimode states~\cite{1367-2630-13-3-033027} with broadband mode operators $A_k$ by 
\begin{eqnarray}
g_h^{(2)}(0) = \frac{\left\langle\left(\sum_{j,m}A_j^\dagger A_m^\dagger A_jA_m\right)\right\rangle}{\left\langle \sum_j A_k^\dagger A\right\rangle^2}.
\end{eqnarray}
Substituting $\rho_s$ and neglecting the reflected modes in the denominator results in
\begin{equation}
\fl g_h^{(2)}(0) \approx \frac{2c_{2_t}c_{0_r}p_{herald}\sum_{k_t\le k_t^\prime} (q_{k_t})^2(q_{k_t^\prime})^2}{\left(c_{1_t}c_{0_r}\sum_{k_t} (q_{k_t})^2\right)^2} + \frac{2c_{1_t}c_{1_r}p_{herald}\sum_{k_t} (q_{k_t})^2\sum_{k_r} (q_{k_r})^2}{\left(c_{1_t}c_{0_r}\sum_{k_t} (q_{k_t})^2\right)^2}.
\end{equation}

The factor 2 comes from the annihilation operators themselves when they act on the same mode (i.e. $j=m=k_t=k_t^\prime$), and from the two equivalent arrangements of annihilation operators when they act on different modes (e.g. $j=k_t$, $m=k_r$; or $j=k_r$, $m=k_t$).
We then keep $p_{herald}$ only to first order and assume negligible dark counts such that $c_{0_r}\rightarrow1$, allowing to identify the second term of our approximate fidelity with \gtwo$/2$, resulting in the same fidelity as without filtering, namely
\begin{equation}
F \approx \sqrt{P}\left(1-\frac{g_h^{(2)}(0)}{2}\right).
\end{equation}

\bibliographystyle{iopart-num}
\bibliography{ffwd_bib}
\end{document}